\documentclass[12pt,a4paper]{article}
\usepackage{a4wide}
\usepackage{latexsym}
\usepackage{epsf}
\usepackage{amssymb}

\def\a{\alpha}
\def\b{\beta}
\def\g{\gamma}

\def\m{\mu}
\def\n{\nu}

\def\s{\sigma}
\def\e{\epsilon}

\def\be{\begin{equation}}
\def\ee{\end{equation}}
\def\bea{\begin{eqnarray}}
\def\eea{\end{eqnarray}}

\begin{document}
\begin{flushright}
IFT-UAM/CSIC-03-05\\
hep-th/0302096\\
\end{flushright}

\vspace{1cm}

\begin{center}

{\bf\Large A note on the dyonic D6-brane \footnote{Contribution to the 6th International Workshop on Conformal Field Theory and Integrable Models, Landau Institute Sept. 2002 to honour A. Belavin on the occasion of his 60th birthday.  }     }

\vspace{.7cm}

{\bf C\'esar G\'omez
\footnote{E-mail: {\tt cesar.gomez@uam.es}} and Juan Jos\'e Manjar\'{\i}n}
\footnote{E-mail: {\tt juanjose.manjarin@uam.es}} \\

\vspace{1cm}

{\it   
 Instituto de F\'{\i}sica Te\'orica, C-XVI,
  Universidad Aut\'onoma de Madrid \\
  E-28049-Madrid, Spain}\footnote{Unidad de Investigaci\'on Asociada
  al Centro de F\'{\i}sica Miguel Catal\'an (C.S.I.C.)}

\vskip 1.8cm


{\bf Abstract}
\end{center}
We study the dyon electric charge of D6 branes as eleven dimensional KK monopoles. We observe that the dyon charge is intimately related with the existence of  gauge connections and antisymmetric fields on the brane world volume.

\newpage
\section{Introduction}

The dynamics of open tachyon condensation \cite{sen} and the K-theoretical classification of D-brane charges \cite{mm,witten2}, strongly suggest the interpretation of D-branes as solitons of some ten dimensional underlying gauge-Higgs theory naturaly associated with configurations of D9-filling-branes with the open tachyon playing the role of the Higgs field. 

A fascinating issue, that is still open, is to unravel the precise physical meaning of the gauge theory used in the K-theoretical description of D-brane charges. In this approach, for instance, the D6-brane of type IIA string theory can be naturally interpreted as a 't Hooft-Polyakov monopole \cite{hora,go.ma}. The pattern of symmetry breaking can be characterized by the vacuum manifold $\frac{U(2)}{U(1) \otimes U(1)}$ and the topological stability of the D6-brane is related to the non-vanishing second homotopy group of this vacuum manifold. 

On the other hand, it is well known that 't Hooft-Polyakov monopoles in the standard Yang-Mill-Higgs model in four dimensions have electric charge, i.e are dyons. This is manifest in the topology of the corresponding moduli space that is, for just one monopole, $R^{3} \otimes S^{1}$ \cite{ah}. Of course, it is tempting, but probably too naive, to try to see in the $S^{1}$ of the monopole moduli space a manifestation of the extra eleven dimension of M-Theory. However what we can certainly expect, if the open tachyon condensation approach to D-branes as solitons is correct, is to find some dyon charge for the D6-brane. 

The interesting thing about D6-branes is that they have a purely gravitational description as KK-monopoles in eleven dimensional supergravity \cite{town,hull}.  It is well known that KK-monopoles in ten dimensions have an electric dyon charge \cite{sen2}. Moreover, the moduli of these ten dimensional KK-monopoles is in fact $R^{3} \otimes S^{1}$ as it is the case for the dyon. The extra $S^{1}$ part of the moduli in this case is related with large gauge transformations of the $B$-field and the electric charge of the KK-monopole can be interpreted as a consequence of unwinding the string modes on a Hopf fibration of $S^{3}$ \cite{ghm}. In this sense what we need to do in order to prove that D6-branes are dyons, as it is predicted by the open tachyon condensation description of them, is to extend the previous argument on ten dimensional KK-monopoles to the case of eleven dimensional KK-monopoles. This is the modest target of this note.

\section{Eleven Dimensional Description of D6-branes}\label{s2}

Let us start recalling some well known facts about D6-branes as eleven dimensional KK-monopoles

At the level of solutions, the D6-brane has a singularity at the origin which makes necessary the eleven dimensional description \cite{town} as a KK-monopole, in order to have a completely regular solution. We can check this explicitely by means of the usual KK reduction for the eleven dimensional metric

\be
\frac{ds^2_{11}}{l_p^2}=e^{4\phi/3}\left( dx^s+{\cal A}_\m dx^\m\right)^2+\frac{e^{-2\phi/3}}{l_s^2}ds^2_{10}.
\ee

\noindent The relations between the M-theory parameters $(l_p,R_s)$ and the string theory parameters $(l_s,g_s)$ are

\bea
g_s=e^{2\phi},\quad \frac{R_s}{l_p}=g_s^{2/3},\quad l_p^3=g_sl_s^3.
\eea

When considering extended objects, we can decompose the metric as ${\mathbb E}^{1+p}+{\mathbb E}^{10-p}$. The solution is then determined by an harmonic function, $H$, in the transverse space

\be
H=1+\frac{k}{r^{8-p}}.
\ee

As we are dealing with the D6-brane of type IIA string theory, we can associate it to a 6-form in ten dimensions. However, when this form is lift to eleven dimensions, we find that it is only non-trivial in the presence of an isometric direction, i.e. in the presence of a KK-monopole of the eleven dimensional theory \cite{hull}. 

The metric for the KK-monopole can be written as

\be
ds^2_{11}=Hdx^\m dx_\m+H^{-1}\left( d\psi_{TN}+{\cal A}_\m dx^\m\right)^2+ds^2\left({\mathbb E}^{1,6}\right),
\ee

\noindent where $\psi_{TN}$ is taken as a periodic coordinate of period $2\pi$ (the Taub-NUT coordinate) in order to eliminate the Dirac string singularity and which will represent the compact eleventh dimension. For $G=0$ (with $G$ the field strenght of the 3-form of M-theory), $\nabla\wedge{\cal A}=\nabla H$, which implies $\nabla^2{\cal A}=0$.

It is worth to mention that this solution is a gravitational instanton of the eleven dimensional supergravity, of the {\sl nut} type, which interpolates between the KK-vaccum ${\mathbb E}^{1,9}\times S^1$ at $r\rightarrow\infty$ and the M-theory vacuum at $r\rightarrow 0$.

Performing a direct dimensional reduction we find the solution for the D6-brane

\be
ds_{10}^2=H^{-1/2}ds^2({\mathbb E}^{1,6})+H^{1/2}ds^2({\mathbb E}^{3}),
\ee

\noindent with

\bea
e^{-2\phi}=H^{3/2},\quad F_m={^*}dH\quad{\mbox{and }}H=1+\frac{k}{r},
\eea

\noindent where $F_m$ is the magnetic field strenght and the Hodge dual is taken in the transverse space.

This process of desingularization has also the advantage of making evident the origin of the field content of the D6-brane, which is clear from the point of view of the action principles \cite{bjo}. As usually stated, the field content of the KK-monopole is a seven dimensional vector multiplet of $N=1$ supersymmetry consisting on the vector field $A_i$ and three scalars corresponding to transaltional zero modes in the transverse space.  

Apart from the purely gravitational part, this field constent arises from the dimensional reduction of the M2-brane. More especifically, we can consider the decomposition of the $C^{(3)}$ form, to which the eleven dimensional membrane couples, in the presence of the KK-monopole as

\be\label{C}
C^{(3)}_{MNP}\stackrel{KKM}{\longrightarrow}\left\{\matrix{C_{\m\n\rho} & & \cr C_{\m\n i} & = & C_{\m\n}A_i \cr C_{\m ij} & = & V_\m B_{ij} \cr C_{ijk}}\right. ,
\ee

\noindent where $x^\m=(r,\theta,\phi,\psi)$ are the Taub-NUT coordinates (with $\psi\equiv x^\#$ the eleventh dimensional coordinate) and $y^i$, $i=0,...,6$ the coordinates on the world-volume of the KK-monopole. The fields $A$ and $B$ are respectively the gauge field and the antisymmetric 2-form on the D6-brane world volume. 

As is clear from the field content of the KK-monopole, this decomposition implies that the system we are considering is an M2-brane wrapped on the isometric direction of the space with an end on the world-volume of the KK-monopole, which can be schematically represented as

\be
(0|M2,KKM)=\left\{\matrix{\times & | & - & - & \times & \times & - & - & - & - & - & - \cr
                                                          \times & | & - & - & - & z & \times & \times & \times & \times & \times & \times}\right. .
\ee

The zero modes in the decomposition of the 3-form of M-theory which live in the transverse space will be determined by the geometry of the space, in the sense that they will be requiered to be harmonic functions

\bea
\nabla^\a\nabla_{[\a}V_{\m]}=0,\\
\nabla^\a\nabla_{[\a}C_{\m\n]}=0,
\eea

\noindent which, after a proper parametrization to be explained below, implies \cite{ima}

\be\label{G}
\nabla_{[\m}V_{\n]}=\frac{1}{2\pi\a'}C_{\m\n},
\ee

\noindent which can be seen as the gauge transformation for $C_{\m\n}$ and establishes the standard relation between the B-field and the gauge field on the D6-brane, as can be seen building the corresponding part of the 4-form field strenght as

\bea
\label{gm}
\nonumber G_{\m\n ij} & =& C_{\m\n}\nabla_{[i}A_{j]}+\nabla_{[\m}A_{\n]}B_{ij}\\
\nonumber & = & C_{\m\n}\left( \nabla_{[i}A_{j]}+\frac{1}{2\pi\a'}B_{ij}\right)\\
& = & C_{\m\n}{\cal F}_{ij}.
\eea 

\noindent When the $C_{\m\n}$ is integrated, gives rise to the $F+B$ term of the Born-Infeld theory of the D6-brane.

Moreover, due to the holonomy group of the Taub-NUT manifold, in order to obtain a normalizable $C_{\m\n}$, we must require for it to be an anti-self-dual tensor.

Let us stop for a second and look at the implications of (\ref{G}). We have obtained a description of the KK-monopole from a purely gravitational point of view, including the degrees of freedom corresponding to the gauge fields of the N=1 vector multiplet of its world-volume theory, as seen from (\ref{gm}).

There is, however, an extra condition which we have not used, namely, that the field strenght of the 3-form of M-theory, $G$, has to be zero. From (\ref{gm}) one obtains the following relation between the fields on the world-volume

\be
\label{F}
\frac{1}{2\pi\a'}B+dA=0
\ee 

\noindent Which trivially implies that the B-field is flat, i.e. its topological classification lies in $[H]\in Tors\left( H^3(X,{\mathbb{Z}})\right)$.

\section{D6-brane Charges}\label{s3}

Now we can compute the charges of the system. The metric on the transverse space can be written as

\be
\label{tnm}
ds^2_{TN}=\frac{r+m}{r-m}dr^2+(r^2-m^2)d\Omega^2_{(2)}+(4m)^2\frac{r-m}{r+m}\left(d\psi+\frac{\cos\theta}{2}d\phi\right)^2,
\ee

\noindent so the vector potential and its field strenght are

\bea
A&=&2m\cos\theta d\phi,\\
F&=&dA=-2m\sin\theta d\theta\wedge d\phi.
\eea

From here one can compute the NUT charge as 

\be
N=\frac{1}{8\pi}\int F=m.
\ee

It is interesting to mention that the parameter $m$ is set, from boundary conditions to be

\be
4m=l_se^{\phi_0}
\ee

\noindent where $\phi_0$ is the boundary value of the dilaton of the type IIA string theory.

The magnetic charge of the monopole can be computed from the integral of the totally antisymmetric part of the spin connection minus the background $\omega=F\wedge k$, where $k=\frac{r-m}{r+m}\left(4md\psi+2m\cos\theta d\phi\right)$ is the Killing vector representing the isometry, which is, up to a normalization factor

\be
\label{mf}
\omega=-(4m)^2\frac{r-m}{r+m}\frac{\sin\theta}{2}d\theta\wedge d\phi\wedge d\psi.
\ee

\noindent This form is proportional to the volume form of the $S^3$ which means that it can be interpreted as a winding number. The corresponding charge is

\be
K=\frac{1}{16\pi^2}\int \omega=4m^2.
\ee

The metric (\ref{tnm}) is self-dual, i.e. $R_{\a\b\g\s}=\frac{1}{2}\e_{\a\b\eta\pi}R^{\eta\pi}{ }_{\g\s}$, which, in turn, implies that the mass of the solution is equal to its NUT charge, and has a positive orientation defined by an orthonormal frame \cite{gh} which can be chosen in such a way that

\bea
e^\psi & = & \left(\frac{r-m}{r+m}\right)^{1/2}\left(4md\psi+2m\cos\theta d\phi\right),\\
e^r       & = & \left(\frac{r+m}{r-m}\right)^{1/2}dr,
\eea

\noindent where we have just written the two components that will be interesting for us. 

At this moment we have computed three of the charges of the system, and this would be all if we do not consider the vector potentials associated with the 3-form of M-theory.

In order to define now the electric charge we will use the following strategy. Since we are considering an eleven dimensional supergravity solution, we will use the M-theory 3-form $C_{MNP}$. The pieces of $C_{MNP}$ that will be relevant for our discussion are the 2-form $C_{\m\n}$ and the 1-form $V_{\m}$ defined in (\ref{C}). Notice that these forms are associated, respectively, with the gauge field $A_{i}$ and the 2-form $B_{ij}$ on the D6-brane world-volume.

Using (\ref{G}), we observe that the 2-form $C_{\m\n}$ is a pure gauge. It is this pure gauge 2-form the one that is playing for the eleven dimensional KK-monopole case the same role that a pure gauge 2-form $B$ in the case of the ten dimensional KK-monopole. In this sense we will define the electric charge as associated with $C_{\m 4}$ where the coordinate $4$ is the one of the $S^{1}$ fiber of the Hopf fibration. 

Let us stress that the main difference with the dyon effect for the ten dimensional KK-monopole is that we need to proyect the M-theory 3-form $C$ on the world-volume coordinates. Moreover, the fact that the so defined 2-form is a pure gauge is reflecting the gauge invariance of the world-volume BI lagrangian. 

Thus in order to compute the electric charge, we parametrize the $V_\m$ as in \cite{ima}

\be
V_\m=\left( f_1(r),0,\frac{1}{2}f_2(r)\cos\theta,f_2(r)\right),
\ee

\noindent which can be seen to correspond, up to normalization constants, to the vector $k$ plus a deformation coming from the $e^r$. This allows us to write

\be
C_{\m 4}=-\frac{1}{(2\pi)^3l_s^{3/2}}\frac{2m}{(r+m)^2}dr.
\ee

Dualising this form we find

\be
\label{ef}
C_{(3)}={^*}C_{\m 4}=-\frac{3(4m)^2}{2(2\pi)^3l_s^{3/2}}\left(\frac{r-m}{r+m}\right)^2\sin\theta d\theta\wedge d\phi\wedge d\psi,
\ee

\noindent which is again proportional to the volume of the $S^3$. This is the electric charge of the KK-monopole, and is well defined at infinity.

Notice that the $S^{1}$ part of the dyon moduli space is related to large gauge transformations of the 2-form $C_{\m\n}$.

This concludes the proof that the D6-brane interpreted as an eleven dimensional KK-monopole is -as predicted from the ten dimensional open tachyon condensation description-  a dyon. This points out to a deep connection between the gauge-Higgs ten dimensional description of a D6-brane as a  't Hooft-Polyakov monopole and the eleven dimensional  gravitational description. The dyon charge is the most natural bridge between the two descriptions \cite{ces}.

\section*{Acknowledgements}

This research was possible thanks to the grant AEN2000-1584.


\end{document}